\documentclass[10pt]{article} 
\usepackage{amssymb}
\usepackage{amsmath}
\usepackage{amsthm}
\usepackage{latexsym} 
\usepackage[dvips]{epsfig}

\theoremstyle{plain}
 
\newtheorem{lemma}{Lemma}
\newtheorem{theorem}{Theorem}

\newtheorem*{main}{Theorem}

\def\O{\mathcal{O}}

\def\xt{\widetilde{x}}

\font\tenscr=rsfs10 scaled1100
\font\sevenscr=rsfs7 
\font\fivescr=rsfs5 
\skewchar\tenscr='177 
\skewchar\sevenscr='177
\skewchar\fivescr='177 
\newfam\scrfam 
\textfont\scrfam=\tenscr
\scriptfont\scrfam=\sevenscr 
\scriptscriptfont\scrfam=\fivescr

\begin{document}


\title{Asymptotic expansions of the Cotton-York tensor on slices of stationary spacetimes}

\author{Juan Antonio Valiente Kroon \thanks{E-mail address:
 {\tt jav@ap.univie.ac.at}} \\
 Institut f\"ur Theoretische Physik,\\ Universit\"at Wien,\\
Boltzmanngasse 5, A-1090 Wien,\\ Austria.}

\maketitle

\begin{abstract}
We discuss expansions for the Cotton-York tensor near infinity for
arbitrary slices of stationary spacetimes. From these expansions it
follows directly that a necessary condition for the existence of
conformally flat slices in stationary solutions is the vanishing of a
certain quantity of quadrupolar nature (obstruction). The obstruction
is nonzero for the Kerr solution. Thus, the Kerr metric admits no
conformally flat slices. An analysis of the next order terms in the
expansions in the case of  solutions such that the
obstruction vanishes, suggests that the only stationary solutions
admitting conformally flat slices are the Schwarzschild family of solutions.
\end{abstract}


\section{Introduction}

The question whether the Kerr spacetime admits a Cauchy hypersurface
that is conformally flat is an issue which has been of particular
interest for the numerical relativity community. The reason for this
interest lies in the extensive use of conformally flat initial data
sets ---like the Bowen-York and Brandt-Br\"ugmann data
\cite{BowYor80,BraBru97}--- in the numerical simulations of the
collision of, say, two spinning black holes. Therefore, it would be
very convenient to have a family of conformally flat data sets which
would reduce to a Kerr initial data set under a particular choice of
the parameters of the family. Such a family of data would then allow
to study the close limit case as perturbations of a Kerr black hole
in, for example, a way analogous to the work carried out by Gleiser
\emph{et al} \cite{GleNicPriPul98}. In their analysis, they considered
the close limit for Bowen-York data ---which does not reduce to Kerr
data under any choice of the parameters. Hence, they were forced to
study the close limit as perturbations of a Schwarzschild black
hole. Accordingly, they were only able to consider situations with
small angular momentum.

The existence or not of such conformally flat slices is also
interesting from a more theoretical point of view. More generally, one
could ask whether a generic asymptotically flat, stationary spacetime
admits conformally flat Cauchy slices or not. For a while now, there
has been the suspicion that there are no such slices ---unless, of
course, the spacetime corresponds to the Minkowski or Schwarzschild
solution.

There are a couple of partial negative answers to the aforementioned
questions. Garat \& Price \cite{GarPri00} have shown that there are no
axially symmetric, conformally flat slices in a given Kerr spacetime
which smoothly reduce to the standard Schwarzschildean $t=constant$
slices in the limit when the angular momentum parameter $a$ goes to
zero. Their approach is what one could call the obvious way of dealing
with the problem. That is, they calculated the Cotton-York tensor of
their slices, and found that they could not be chosen so that the
Cotton-York tensor vanishes \footnote{The Cotton-York tensor,
sometimes also called the Bach tensor, locally characterises the
conformal flatness of 3-dimensional manifolds. The Cotton-York
vanishes if and only if the the 3-dimensional manifold is locally
conformally flat. It can be thought of as the 3-dimensional analogue
of the Weyl tensor. A classical discussion of this fact can be found
in Eisenhart's book, chapter II, section 28 \cite{Eisenhart}. A
discussion in terms of more modern language can be found in
\cite{GarHehHeiMac04,Her97}. As an historical remark, it is noted that
the first use of this object in General Relativity seems to be found
in \cite{ArnDesMis62}, chapter 7, section 7-5.4. I thank S. Deser for
this observation.}. The Cotton-York tensor is a complicated object
involving derivatives of the curvature of the slice. Thus, in order to
keep their calculations tractable, they considered expansions in
powers of the angular momentum parameter $a$. Remarkably, the
\emph{obstructions} to the existence of conformally flat slices arise
at order $a^2$.

More recently, in \cite{Val04b} it was shown that stationary solutions
with non-vanishing angular momentum admit no conformally flat,
maximal, non-boosted (i.e. with vanishing linear momentum) Cauchy
slices. This result was obtained by calculating a certain type of
asymptotic expansions near null and spatial infinity. Notably, the
only fact about stationary solutions used in this result is that they
are known to admit a smooth null infinity.

The results obtained in \cite{GarPri00,Val04b} are not fully
satisfactory because of the assumptions they make about the nature of
the slices. The stationary solutions are well understood, and in a
certain sense ---via their multipolar expansions---, one could say all
is known about them. Thus, it ought to be possible to reduce the
assumptions regarding the potential conformally flat slices to the
bare minimum. This is the goal of the present work. Our analysis shows
that, in order to have conformally flat slices in a given stationary
spacetime, certain quantities of quadrupolar nature ---to which we
shall refer to as \emph{obstructions}--- should vanish. For the
particular case of the Kerr solution these obstructions do not vanish,
whence this spacetime admits no conformally flat Cauchy
hypersurfaces. The only assumption made here about the slices is that
they should have an intrinsic 3-metric for
which a conformal compactification at infinity is well defined. In
order to prove the aforementioned result we have used an approach that
can be regarded as an hybrid of the approaches used in \cite{GarPri00}
and \cite{Val04b}: we have made use of the multipolar expansions for
stationary spacetimes to calculate asymptotic expansions of the
Cotton-York tensor near infinity. By looking at orders in the
expansions higher than the one required to prove our main result, we also
obtain further evidence to the suspicion that the only stationary
spacetimes admitting conformally flat slices are the Schwarzschild and
Minkowski solutions.

In this work we have made extensive use of multipolar expansions
similar to those introduced by Simon \& Beig in reference
\cite{SimBei83}. Given the high order in the expansions needed to
obtain our results, most of the calculations have been performed using
the computer algebra system {\tt Maple V}. An example, based on the
subclass of axially symmetric solutions, of how the computer algebra
implementation was carried out can be found in an accompanying {\tt
Maple V} script. Due to the size of some of these expressions, most of
them are not given in full. The article is organised as follows:
section 2 reviews some of the basic features and formulas concerning
multipolar expansions that have been used in our analysis. In
particular, we discuss the so-called Geroch-Hansen potentials in both
the quotient and rescaled quotient manifolds. The connection of the
metrics of these quotient manifolds to the metrics of Cauchy
hypersurfaces of the spacetime are also considered. Section 3 is
concerned with the asymptotic expansions of the Cotton tensor for some
particular slices which we shall call ``canonic''. In section 4 the
expansions for more generic slices are discussed. Our main result is
presented there. Finally in section 5, there are some comments
concerning higher order expansions.

\medskip
\textbf{Notation.} Greek indices are spacetime ones, while latin
indices are spatial. We have adopted the definition of the Cotton-York
tensor as given in the \emph{Exact Solutions} book \cite{KSMH}. The
convention used for spherical harmonics is the one given in Arfken's
book \cite{WebArf00}. A quantity $\phi(x^i)$ will be said to be $\O^\infty(f(r))$ if there is a $C^\infty$ function $f(r)$ such that $|\phi|\leq |f|$, $\partial_i\phi|\leq |\partial f/\partial r|$, $|\partial_i \partial_j \phi|\leq |\partial^2f/\partial r^2|,\cdots$.
 
\section{Stationary spacetimes and their multipolar expansions}

Let $(\widetilde{M},\widetilde{g}_{\mu\nu})$ be a stationary vacuum
spacetime, and let $\xi^\mu$ be the corresponding timelike Killing
vector. The collection of the orbits of the Killing vector $\xi^\mu$
defines the so-called \emph{quotient manifold}, $\widetilde{X}$. In
terms of quantities defined on the quotient manifold, the metric
$\widetilde{g}_{\mu\nu}$ locally takes the form
\begin{equation}
\label{metric:quotient}
\widetilde{g}=\lambda\left(dt+\beta_i d\widetilde{x}^i\right)^2-\lambda^{-1}\widetilde{\gamma}_{ij}d\widetilde{x}^id\widetilde{x}^j, \qquad i,j=1,2,3,
\end{equation} 
where $\lambda$, $\beta_i$ and $\widetilde{\gamma}_{ij}$ \emph{(the
metric of the quotient manifold)} depend only on the spatial
coordinates $\widetilde{x}^k$. For the purposes of our later
discussion, we will also need to have the metric $\widetilde{g}$
written in terms of a 3+1 decomposition with respect to the
 $t=constant$ hypersurfaces. One has
\begin{equation}
\label{metric:adm}
\widetilde{g}=\widetilde{N}^2dt^2-\widetilde{h}_{ij}\left(N^idt+d\widetilde{x}^i\right)\left(N^jdt+d\widetilde{x}^j\right).
\end{equation}

The relation between the two sets of quantities is given by ---see e.g. \cite{Dai01b}---
\begin{eqnarray}
&& \widetilde{N}^2=\frac{\lambda}{1-\lambda^2\widetilde{\beta}^i\beta_i}, \\
&& N_j=\lambda \beta_j, \qquad N^j=-\widetilde{N}^2\lambda \widetilde{\beta}^j, \\
&& \widetilde{h}_{ij}=\lambda^{-1}\widetilde{\gamma}_{ij}-\lambda\beta_i\beta_j, \qquad \widetilde{h}^{ij}=\lambda\widetilde{\gamma}^{ij}+\lambda^2\widetilde{N}^2\beta^i\beta^j,
\end{eqnarray}
where
\begin{equation}
\widetilde{N}_j=\widetilde{h}_{ij}N^i, \qquad \widetilde{\beta}^j=\widetilde{\gamma}^{jk}\beta_k.
\end{equation}

The stationary field equations can be conveniently be written in terms
of certain quantities defined in the quotient manifold: the metric
$\widetilde{\gamma}_{ij}$, and 3 potentials, $\widetilde{\phi}_M$,
$\widetilde{\phi}_S$, $\widetilde{\phi}_K$. These potentials are
certain algebraic combinations of
$\lambda=\widetilde{g}_{\mu\nu}\xi^\mu\xi^\nu$ (the square of the norm
of the Killing vector field) and of a scalar $\omega$ (the twist of
$\xi^\mu$) such that
\begin{equation}
\label{twist}
\widetilde{D}_i\omega=\omega_i=-\lambda^2\sqrt{\det \widetilde{\gamma}}\epsilon_{ijk}\widetilde{D}^j\beta^k.
\end{equation}
More precisely, one defines
\begin{equation}
\widetilde{\phi}_M=\frac{1}{4\lambda}(\lambda^2+\omega^2-1), \quad \widetilde{\phi}_S=\frac{1}{2\lambda}\omega, \quad \widetilde{\phi}_K=\frac{1}{4\lambda}(\lambda^2+\omega^2+1).
\end{equation}
These potentials are not independent, but satisfy the relation
$\widetilde{\phi}_M^2+\widetilde{\phi}_S^2-\widetilde{\phi}_K^2=-1/4$. The
stationary vacuum field equations read:
\begin{subequations}
\begin{eqnarray}
&& \widetilde{D}^i\widetilde{D}_i \widetilde{\phi}=2\widetilde{R}\widetilde{\phi}, \label{stationary_1}\\
&& \widetilde{R}_{ij}=2\left(\widetilde{D}_i\widetilde{\phi}_M\widetilde{D}_j\widetilde{\phi}_M+\widetilde{D}_i\widetilde{\phi}_S\widetilde{D}_j\widetilde{\phi}_S-\widetilde{D}_i\widetilde{\phi}_K\widetilde{D}_j\widetilde{\phi}_K\right), \label{stationary_2}
\end{eqnarray}
\end{subequations}
where $\widetilde{\phi}$ denotes any of the potentials
$\widetilde{\phi}_M$, $\widetilde{\phi}_S$, $\widetilde{\phi}_K$ and
$\widetilde{R}_{ij}$ is the Ricci tensor of $\widetilde{\gamma}_{ij}$, and $\widetilde{R}$ its Ricci scalar.

\subsection{Expansions in the quotient manifold $\widetilde{X}$}

Simon \& Beig \cite{SimBei83} have studied certain (asymptotic)
expansions of the potentials $\widetilde{\phi}_M$,
$\widetilde{\phi}_S$, $\widetilde{\phi}_K$ which are closely related
to the Geroch-Hansen multipoles \cite{Han74}. More precisely, they
have shown that:

\begin{theorem}[Simon \& Beig, 1983] \label{SimonBeig}
For all stationary vacuum solutions there is a (Cartesian) coordinate
system $\xt^i$ ($\widetilde{r}=\sqrt{\delta_{ij}\xt^i\xt^j}$) on the
quotient manifold $\widetilde{X}$ and there are sets of constants
$A_{\cdots}$, $B_{\cdots}$, $G_{\cdots}$ such that for all nonnegative
integers $m$
\begin{eqnarray*}
&&\widetilde{\phi}_M=\sum^{m-1}_{l=0}\frac{E_{a_1\cdots a_l}\xt^{a_1}\cdots \xt^{a_l}}{l! \widetilde{r}^{2l+1}}+O^\infty(\widetilde{r}^{-(m+1)}), \\
&&\widetilde{\phi}_S=\sum^{m-1}_{l=1}\frac{F_{a_1\cdots a_l}\xt^{a_1}\cdots \xt^{a_l}}{l! \widetilde{r}^{2l+1}}+O^\infty(\widetilde{r}^{-(m+1)}), \\
&&\widetilde{\phi}_K=\frac{1}{2}+\sum^{m-1}_{l=0}\frac{G_{a_1\cdots a_l}\xt^{a_1}\cdots \xt^{a_{l-1}}}{l! \widetilde{r}^{2l+1}}+O^\infty(\widetilde{r}^{-(m+1)}), \\
&&\widetilde{\gamma}_{ij}=\delta_{ij}+\sum^m_{l=2} \left( \frac{\xt_i \xt_j A_{a_1\cdots a_{l-2}} \xt^{a_1}\cdots \xt^{a_{l-2}}}{\widetilde{r}^{2l}}+\frac{\delta_{ij}B_{a_1\cdots a_{l-2}} \xt^{a_1}\cdots \xt^{a_{l-2}} }{\widetilde{r}^{2l-2}} \right. \\
&& \phantom{XXXXXXXXX}\left. + \frac{\xt_{(i}C_{j)a_1\cdots a_{l-3}} \xt^{a_1}\cdots \xt^{a_{l-3}}}{\widetilde{r}^{2l-2}} + \frac{D_{ija_1\cdots a_{l-4}} \xt^{a_1}\cdots \xt^{a_{l-4}}}{\widetilde{r}^{2l-4}}  \right) +\O^\infty(\widetilde{r}^{-(m+1)}).
\end{eqnarray*}
All constants are symmetric in all their $a_j$ indices. $A_{a_1\cdots a_{l-2}}$, $B_{a_1\cdots a_{l-2}}$, $C_{ja_1\cdots a_{l-3}}$, $D_{ija_1\cdots a_{l-4}}$, $G_{a_1\cdots a_{l-1}}$ and the trace parts of $E_{a_1\cdots a_{l-1}}$ and of $F_{a_1\cdots a_{l-1}}$ for $0\leq l \leq m$, depend on the trace free parts of $E_{a_1\cdots a_{l'-1}}$ and $F_{a_1\cdots a_{l'-1}}$, with $1\leq l'\leq m$. 
\end{theorem}

In \cite{SimBei83} the explicit form of these expansions for $m=3$ has
been calculated. Unfortunately, it turns out that our analysis
actually requires knowing the expansions up to $m=5$. For
computational purposes ---as it is more amenable to a computer algebra
implementation--- it is more convenient to work not in the Cartesian
coordinates discussed in the aforementioned theorem, but in the
associated spherical coordinates given by
\[
\widetilde{x}^1=\widetilde{r} \sin \theta \cos \varphi, \quad \widetilde{x}^2=\widetilde{r} \sin \theta \sin \varphi, \quad \widetilde{x}^3=\widetilde{r}\cos\theta.
\]

Based on theorem \ref{SimonBeig}, we make the following Ansatz for the
expansions (in spherical coordinates) of the Hansen potentials and the
metric of the quotient manifold:

\begin{subequations}
\begin{eqnarray}
&&\widetilde{\phi}_M=\frac{M}{\widetilde{r}}+\sum_{k=3}^5\sum_{l=0}^{k-1} \sum_{m=-l}^l \frac{M_{klm}Y_{lm}}{\widetilde{r}^k}+\O\left(\frac{1}{\widetilde{r}^6}\right), \\
&&\widetilde{\phi}_S=\frac{S_{200}Y_{00}+S_{210}Y_{10}}{\widetilde{r}^2}+\sum_{k=3}^5\sum_{l=0}^{k-1} \sum_{m=-l}^l \frac{S_{klm}Y_{lm}}{\widetilde{r}^k}+\O\left(\frac{1}{\widetilde{r}^6}\right), \\
&&\widetilde{\phi}_K=\frac{1}{2}+\sum_{k=2}^5\sum_{l=0}^{k-1} \sum_{m=-l}^l \frac{K_{klm}Y_{lm}}{\widetilde{r}^k}+\O\left(\frac{1}{\widetilde{r}^6}\right),
\end{eqnarray}
\end{subequations}
where $M(\neq0)$ denotes the mass of the stationary spacetime, and $Y_{lm}$
are standard spherical harmonics expressed in terms of spherical
coordinates. If the stationary spacetime is also axially symmetric the
above expressions contain only $Y_{l0}$ harmonics. In the above
expansions the angular momentum monopole has been set to zero in order
to guarantee asymptotic flatness. Furthermore, a translation and a
rotation have been used ---without loss of generality--- to set the
mass dipolar terms equal to zero and to ``align the angular momentum
along the z axis'', that is, to set equal to zero the coefficients
coming with the spherical harmonics $Y_{11}$ and $Y_{1-1}$ at order
$1/\widetilde{r}^2$ in $\widetilde{\phi}_S$. The potentials
$\widetilde{\phi}_M$, $\widetilde{\phi}_S$ and $\widetilde{\phi}_K$
are real. Hence, the diverse coefficients in the expansions have to
satisfy the reality conditions
\begin{equation}
M_{kl-m}=(-1)^m\overline{M}_{klm}, \quad S_{kl-m}=(-1)^m\overline{S}_{klm}, \quad K_{kl-m}=(-1)^m\overline{K}_{klm}.
\end{equation}
Similarly, we write for the components of the metric $\widetilde{\gamma}_{ij}$,
\begin{subequations}
\begin{eqnarray}
&& \widetilde{\gamma}_{\widetilde{r}\widetilde{r}}=1+\sum_{k=2}^4\sum_{l=0}^{k-1} \sum_{m=-l}^l \frac{A_{klm}Y_{lm}}{\widetilde{r}^k}+\O\left(\frac{1}{\widetilde{r}^5}\right), \\
&& \widetilde{\gamma}_{\widetilde{r}\theta}=\sum_{k=1}^3\sum_{l=0}^{k-1} \sum_{m=-l}^l \frac{B_{klm}Y_{lm}}{\widetilde{r}^k}+\O\left(\frac{1}{\widetilde{r}^4}\right), \\
&& \widetilde{\gamma}_{\widetilde{r}\varphi}=\sum_{k=1}^3\sum_{l=0}^{k-1} \sum_{m=-l}^l \frac{C_{klm}\sin\theta Y_{lm}}{\widetilde{r}^k}+\O\left(\frac{1}{\widetilde{r}^4}\right), \\
&& \widetilde{\gamma}_{\theta\theta}=\widetilde{r}^2+\sum_{k=0}^2\sum_{l=0}^{k-1} \sum_{m=-l}^l \frac{D_{klm} Y_{lm}}{\widetilde{r}^k}+\O\left(\frac{1}{\widetilde{r}^3}\right), \\
&& \widetilde{\gamma}_{\theta\varphi}=\sum_{k=0}^2\sum_{l=0}^{k-1} \sum_{m=-l}^l \frac{E_{klm}\sin\theta Y_{lm}}{\widetilde{r}^k}+\O\left(\frac{1}{\widetilde{r}^3}\right), \\
&& \widetilde{\gamma}_{\varphi\varphi}=\widetilde{r}^2\sin^2\theta+\sum_{k=0}^2\sum_{l=0}^{k-1} \sum_{m=-l}^l \frac{F_{klm}\sin^2\theta Y_{lm}}{\widetilde{r}^k}+\O\left(\frac{1}{\widetilde{r}^3}\right).
\end{eqnarray}
\end{subequations}
Again, the different coefficients in the latter expansions are
subject to reality conditions analogous to those for the potentials.

The substitution of the Ansatz for the Hansen potentials and the metric of the quotient manifold into the stationary vacuum field equations (\ref{stationary_1}) and (\ref{stationary_2}) yields ---consistently with theorem \ref{SimonBeig}--- that:
\begin{itemize}
\item[(i)] the coefficients $S_{200}$, $M_{klm}$, $S_{klm}$ with
$k=3,\ldots,5$, $l=0,\ldots, k-2$, $m=-l,\ldots,l$ in
$\widetilde{\phi}_M$ and $\widetilde{\phi}_S$;
\item[(ii)] all the coefficients $K_{klm}$ in $\widetilde{\phi}_K$;
\item[(iii)] and all the coefficients $A_{klm}$, $B_{klm}$, $C_{klm}$,
$D_{klm}$, $E_{klm}$, $F_{klm}$ in the components of
$\widetilde{\gamma}_{ij}$,
 \end{itemize}
can be written in terms of the coefficients $M$, $M_{k,k-1,m}$ and
$S_{k,k-1,m}$, $m=1-k,\ldots,k-1$. These coefficients are essentially
the multipole moments of Geroch \& Hansen. The explicit form
dependence in these expansions for the axially symmetric case can be
found in the accompanying {\tt Maple V} script.

\medskip
\textbf{The Kerr solution.} The Kerr spacetime is, arguably, the most
important stationary solution. Part of its relevance lies in the fact
that stationary solutions are Kerrian at first order in the angular
momentum ---see e.g. \cite{BeiSim80}. A quick calculation with the
Kerr initial data in Bowen-York coordinates reveals that
\begin{equation}
\label{kerr:multipoles}
S_{110}=2\sqrt{\frac{\pi}{3}}Ma, \qquad Q_{200}=-86\sqrt{\frac{\pi}{5}}Ma^2,
\end{equation}
 where $a$ is the Kerr parameter ---which has dimensions of angular
 momentum per unit mass. Furthermore, because of the axial symmetry
\begin{equation}
M_{222}=M_{221}=0.
\end{equation}

\subsection{Expansions in the conformally rescaled quotient manifold $X$}
On calculational grounds, it will be much more convenient for us to
work not in $\widetilde{X}$ but in a conveniently conformally rescaled
version thereof. The asymptotic flatness of the stationary spacetime
ensures that there exists a manifold $X$ consisting of $\widetilde{X}$
plus an additional point $i$ such that the metric in $X$ is given by
\begin{equation}
\gamma_{ij}=\Omega^2\widetilde{\gamma}_{ij},
\end{equation}
where 
\begin{equation}
\Omega=\frac{1}{2B^2}\left( \sqrt{1+4(\widetilde{\phi}^2_M+\widetilde{\phi}^2_S)}-1\right),
\end{equation}
for some real constant $B^2>0$. Furthermore, one has that
\begin{equation}
\Omega(i)=0, \qquad D_i\Omega(i)=0 \qquad D_jD_k\Omega(i)=2\gamma_{jk}(i).
\end{equation}
We define the rescaled potentials
\begin{equation}
\phi_M=\widetilde{\phi}_M/\sqrt{\Omega}, \quad \phi_S=\widetilde{\phi}_S/\sqrt{\Omega}, \quad \phi_K=\widetilde{\phi}_K/\sqrt{\Omega}.
\end{equation} 

The expansions discussed in section 2.1 then imply that
\begin{equation}
\Omega=r^2+\left( \left(M^2+\frac{1}{4\pi M^2}S^2_{110}\right)+\sum_{m=-2}^2\left( \frac{2}{M}M_{22m}+\frac{1}{\sqrt{5\pi}M^2}S^2_{110}\delta_{0m}\right)Y_{2m}   \right)r^4+\O(r^5),
\end{equation}
where we have introduced the new coordinate
\begin{equation}
r=1/\widetilde{r},
\end{equation}
and have set $B=M$. In order to perform the calculations to be
described in the following sections, it is actually necessary to know
the expansions of $\Omega$ up to order $r^6$ ---that is, two orders
more! The leading terms of the rescaled potentials read
\begin{subequations}
\begin{eqnarray}
&& \phi_M=M+\left( \left(\frac{M^3}{2}-\frac{1}{8\pi M}S^2_{110}\right)-\frac{1}{2M\sqrt{5\pi}}S^2_{110}Y_{20} \right)r^2+\O(r^3), \label{pot_1}\\
&& \phi_S=S_{110}Y_{10}r+\left(\sum^2_{m=-2} S_{22m}Y_{2m}\right)r^2 +\O(r^3), \label{pot_2}\\
&& \phi_K=\frac{1}{2r}+\left( \left(\frac{3}{4}M^2-\frac{1}{16\pi M^2}S^2_{110}\right)-\sum^2_{m=-2}\left(\frac{1}{2M}M_{22m}-\frac{1}{4\sqrt{5\pi}M^2}S^2_{110}\delta_{0m}\right)Y_{2m}\right)r \nonumber \\
&& \phantom{XXXXXX}+\O(r^2).
\end{eqnarray} 
\end{subequations}
For the purposes of this article, the expansions for $\phi_M$ and
$\phi_S$ need to be known up to order $r^5$, while that of $\phi_K$
only up to order $r^4$. Similarly, the leading terms of the rescaled
metric $\gamma_{ij}$ read
\begin{subequations}
\begin{eqnarray}
&& \gamma_{11}=1+\left( \left(2M^2+\frac{1}{2\pi M^2}S^2_{110}\right)+ \sum^2_{m=-2}\left(\frac{4}{M}M_{22m}+\frac{2}{\sqrt{5\pi}M^2}S^2_{110}\delta_{0m}\right)Y_{2m}\right)r^2 \nonumber \label{met_1}\\
&& \phantom{XXXXX}+\O(r^3), \\
&& \gamma_{12}=\O(r^6), \\
&& \gamma_{13}=\O(r^6), \\
&& \gamma_{22}=r^2+\left( \left(M^2+\frac{1}{2\pi M^2}S^2_{110}\right) +\sum^2_{m=-2}\left(\frac{4}{M}M_{22m}+\frac{2}{\sqrt{5\pi}M^2}S^2_{110}\delta_{0m} \right)Y_{2m} \right)r^4 \nonumber \\
&& \phantom{XXXXX}+\O(r^5), \\
&& \gamma_{23}=\O(r^6), \\
&& \gamma_{33}=\sin^2\theta r^2+\sin^2\theta \left( \left(M^2+\frac{1}{2\pi M^2}S^2_{110}\right) +\sum^2_{m=-2}\left(\frac{4}{M}M_{22m}+\frac{2}{\sqrt{5\pi}M^2}S^2_{110}\delta_{0m} \right)Y_{2m} \right)r^4 \nonumber \label{met_6}\\
&& \phantom{XXXXX}+\O(r^5), 
\end{eqnarray}
\end{subequations}
Later computations will require knowing of the expansions of
$\gamma_{11}$ up to order $r^4$, and up to order $r^6$ for $\gamma_{22}$ and
$\gamma_{33}$. At this point it is worth making the following remark: due
to the work of Beig \& Simon \cite{BeiSim81} we know that for any
triplet
$(\widetilde{\gamma}_{ij},\widetilde{\phi}_M,\widetilde{\phi}_S)$
solution of the stationary field equations (\ref{stationary_1}) and
(\ref{stationary_2}), there is a (Cartesian) coordinate chart around
$i$ in $X$ such that $\widetilde{\gamma}_{ij}$, $\widetilde{\phi}_M$,
$\widetilde{\phi}_S$ and $\Omega$ are analytic. Thus, by a standard
coordinate transformation there is a chart in spherical coordinates
(not necessarily the one we are using) for which expansions as the
ones used above ---equations (\ref{pot_1}), (\ref{pot_2}) and
(\ref{met_1})-(\ref{met_6})--- are well defined. Hence, we shall not
be concerned with questions of convergence of our expansions.

Starting from these expansions, it is not too hard to obtain
those of the scalar $\lambda$ appearing in the metric given by equation 
(\ref{metric:quotient}) ---the square of the norm of the stationary
Killing vector field. Its leading terms read
\begin{eqnarray}
&&\lambda=\frac{1}{2\sqrt{\Omega}(\phi_K-\phi_M)}, \nonumber \\
&&\phantom{\lambda}=1+2Mr+2M^2r^2+2\left(M^3+
\sum_{m=-2}^2 M_{2,2,m}Y_{2,m}\right)r^3+\O(r^4).
\end{eqnarray} 
The calculations in this article require the computation of the above
expansions up to a further order. More complicated is the calculation
of expansions for the shift $\beta_i$ appearing in the metric
(\ref{metric:quotient}). In order to obtain an equation for $\beta^i$
we proceed as follows. We begin by considering the identity
\begin{equation}
\left(\widetilde{D}^i\widetilde{D}_i-\frac{1}{8}\widetilde{R}\right)\widetilde{\phi}=\Omega^{5/2}\left(D^iD_i-\frac{1}{8}R\right)\phi,
\end{equation}
where again $\widetilde{\phi}$ is any of the Hansen potentials. Thus,
the stationary equation (\ref{stationary_1}) can be written as:
\begin{equation} \label{stat_mod}
\left(D^iD_i-\frac{1}{8}r\right)\phi=\frac{15}{8}\Omega^{-2}\widetilde{R}\phi. 
\end{equation}
Now, consider the equation (\ref{stat_mod}) for $\phi_M$ and for
$\phi_S$, multiply them by $\phi_S$ and $\phi_M$ respectively and take
their difference to obtain $\phi_S
D^iD_i\phi_M-\phi_M D^iD_i\phi_S=0$. From here it follows that the
expression $\phi_S D_i\phi_M-\phi_M D_i\phi_S$ is divergence free. An
analogous argument can be used to extract the same conclusion from
$\phi_M D_i\phi_K-\phi_K D_i\phi_M$ and $\phi_S D_i\phi_K-\phi_K
D_i\phi_S$. Thus it follows from equation (\ref{twist}) that $\beta^i$ satisfies the  ``curl-like'' equation
\begin{equation}
\label{equation:curl}
\epsilon_{ijk}D^j\beta^k=\frac{4}{\sqrt{|\gamma|}}\left(\phi_SD_i\phi_K+\phi_MD_i\phi_S-\phi_KD_i\phi_S-\phi_SD_i\phi_M\right).
\end{equation}
Note that because of its ``curl-like'', its
solutions are defined up to a gradient. The properties of the
solutions to this equation have been discussed in lemma 2.5 in
reference \cite{Dai01b}.

\begin{lemma}[Dain, 2001]
There exists a solution $\beta_i$ of equation (\ref{equation:curl}) which in normal (Cartsian) coordinates has the form
\[
\beta_i=\beta_i^1+\frac{1}{r}\beta^2_i,
\]
where $\beta^1_i$ and $\beta^2_i$ are analytic and $\O(x^i)$.
\end{lemma}

 Based on the latter result, we shall consider solutions to equation
(\ref{equation:curl}) whose leading terms are given by
\begin{subequations}
\begin{eqnarray}
&&\beta_r=2i\sqrt{5}\left(S_{221}Y_{11}-\overline{S}_{221}Y_{1,-1}\right)r+\O(r^2), \label{beta1} \\
&& \beta_\theta =\O(r^4) \label{beta2} \\
&& \beta_\phi=-\sin^2\theta\sqrt{\frac{3}{\pi}}S_{110}r+\O(r^2). \label{beta3}
\end{eqnarray}
\end{subequations}
These last expansions need to be calculated up to order $r^4$.

\subsection{The 3-metric of arbitrary slices}
Consistently with our overall strategy, we shall not be interested in the
metric (\ref{metric:quotient}), but in a conveniently conformally
rescaled version thereof. Following \cite{Dai01b}, we define the
(time independent) conformal factor for the spacetime
\begin{equation}
\widehat{\Omega}=\sqrt{\lambda}\Omega,
\end{equation}
so that $g_{\mu\nu}=\widehat{\Omega}^2\widetilde{g}_{\mu\nu}$ and $h_{ij}=\widehat{\Omega}^2\widetilde{h}_{ij}$. Thus,
\begin{equation}
g=\Omega^2\lambda^2(dt+\beta_idx^i)^2-\gamma_{ij}dx^idx^j.
\end{equation}
In terms of a 3+1 decomposition one would have
\begin{equation}
g=N^2dt^2-h_{ij}(N^idt+dx^i)(N^jdt+dx^j).
\end{equation}
Letting
\begin{equation}
N_j=h_{ij}N^i, \qquad \beta^j=\gamma^{jk}\beta_k,
\end{equation}
one finds the following relations between the two decompositions:
\begin{equation}
N=\widehat{\Omega}\widetilde{N}, \quad N=\frac{\Omega^2\lambda^2}{1-\Omega^2\lambda^2\beta_j\beta^j}, \quad \beta^j=-\frac{N^j}{N^2},
\end{equation}
and most importantly,
\begin{equation}
\label{metric_bar}
h_{ij}=\gamma_{ij}-\Omega^2\lambda^2\beta_i\beta_j.
\end{equation}
The aforementioned 3+1 decomposition together with a 1-form $\beta_i$
whose leading terms are given by equations (\ref{beta1})-(\ref{beta3})
render a foliation of the stationary spacetime that in a sense can be
regarded as ``canonical'' ---see the discussion after theorem 2.6 in
\cite{Dai01b}. For example, the slices $t=constant$ in the Kerr
solution given in Boyer-Lindquist coordinates are an example of such a
\emph{canonical foliation}. Any other foliation can be obtained by
introducing a new time coordinate $\overline{t}$ such that
\begin{equation}
\label{new_slice}
t=\overline{t}-F(r,\theta,\varphi).
\end{equation}
The assumptions being made on $F=F(r,\theta,\varphi)$ will be
described later, but for the time being we shall assume it is at
least of class $C^4$ in the coordinates
$(r,\theta,\varphi)$. Substitution of the latter into
(\ref{new_slice}) renders
\begin{equation}
g=\Omega^2\lambda^2\left(d\overline{t}+[\beta_i-\partial_iF]dx^i\right)^2-\gamma_{ij}dx^idx^j.
\end{equation}
Thus, writing $\overline{\beta}_i= \beta_i-\partial_iF$ one finds
that the corresponding 3-metric, $\overline{h}_{ij}$, associated with
the new foliation is given by
\begin{subequations}
\begin{eqnarray}
&&\overline{h}_{ij}=\gamma_{ij}-\Omega^2\lambda^2\overline{\beta}_i\overline{\beta}_j,\\
&&\phantom{\overline{h}_{ij}}=\gamma_{ij}-\Omega^2\lambda^2\left(\beta_i-\partial_iF\right)\left(\beta_j-\partial_jF\right). 
\end{eqnarray}
\end{subequations}

\section{The Cotton-York tensor in the ``canonical'' slices}
As mentioned in the introduction, our strategy is to discuss the
existence of conformally flat slices in stationary spacetimes by
calculating the Cotton-York tensor of the prospective slices. Now, the
Cotton-York tensor is a complicated object. However, on the other
hand, there is some evidence suggesting that generic stationary
solutions do not admit conformally flat slices. The Cotton-York tensor
is a conformal invariant, thus, if it vanishes for a certain
3-dimensional manifold, then it also vanishes for any other manifold
which is conformally related to the original one. Whence, in order to
establish a \emph{no-go} result, we just need, for example, to show
the non-vanishing of the Cotton-York tensor in a neighbourhood of
infinity, $i$ of the slice. The simplest way of doing the latter is by
means of asymptotic expansions around $i$. As a warming up, we firstly
consider the existence of conformally flat slices in the ``canonic''
foliation ---i.e. that for which $F=0$. Given the 3-metric $h_{ij}$,
its Cotton-York tensor is given by
\begin{equation}
\label{cotton:formula}
B^{ij}=2\epsilon^{ikl}D_l\left(R^{j}_{\phantom{j}k}-\frac{1}{4}\delta^j_{\phantom{j}k}R\right),
\end{equation}
where $R^{j}_{\phantom{j}k}$, $R$, and $D_l$ denote respectively the
mixed Ricci tensor, the Ricci scalar and the covariant derivative with
respect to $h_{ij}$. For the case of the so-called canonical slicing
the leading terms of the initial 3-metric read,
\begin{subequations}
\begin{eqnarray}
&& h_{rr}=1+\left( \left(2M^2+\frac{1}{2\pi M^2}S^2_{110}\right)+ \sum^2_{m=-2}\left(\frac{4}{M}M_{22m}+\frac{2}{\sqrt{5\pi}M^2}S^2_{110}\delta_{0m}\right)Y_{2m}\right)r^2 \nonumber \\
&& \phantom{XXXXX}+\O(r^3), \\
&& h_{r\theta}=\O(r^6), \\
&& h_{r\varphi}=\O(r^6), \\
&& h_{\theta\theta}=r^2+\left( \left(M^2+\frac{1}{2\pi M^2}S^2_{110}\right) +\sum^2_{m=-2}\left(\frac{4}{M}M_{22m}+\frac{2}{\sqrt{5\pi}M^2}S^2_{110}\delta_{0m} \right)Y_{2m} \right)r^4 \nonumber \\
&& \phantom{XXXXX}+\O(r^5), \\
&& h_{\theta\varphi}=\O(r^6), \\
&& h_{\varphi\varphi}=\sin^2\theta r^2+\sin^2\theta \left( \left(M^2+\frac{1}{2\pi M^2}S^2_{110}\right) +\sum^2_{m=-2}\left(\frac{4}{M}M_{22m}+\frac{2}{\sqrt{5\pi}M^2}S^2_{110}\delta_{0m} \right)Y_{2m} \right)r^4 \nonumber \\
&& \phantom{XXXXX}+\O(r^5).
\end{eqnarray}
\end{subequations}
The expansions for $h_{rr}$ have been calculated up to order $r^4$,
while those for $h_{\theta\theta}$ and $h_{\varphi\varphi}$ up to
order $r^6$. From here a lengthy calculation with {\tt Maple V} yields
\begin{subequations}
\begin{eqnarray}
&& B_{rr}=\O(r^4), \label{expansion:cotton1}\\
&& B_{r\theta}=iM\left(2\overline{M}_{222}Y_{2,-2}-2M_{222}Y_{22}-\overline{M}_{221}Y_{2,-1}-M_{221}Y_{21}\right)r^4+\O(r^5), \\
&& B_{r\varphi}=\sin\theta\Bigg( \frac{1}{\sqrt{\pi}}\left(45S_{110}^2+2\sqrt{5\pi}MM_{220}\right)\left(\frac{3\sqrt{7}}{35}Y_{30}-\frac{\sqrt{3}}{5}Y_{1,0}\right) \nonumber \\
&& \phantom{XXX}+M\overline{M}_{221}\left(\frac{3\sqrt{5}}{5}Y_{1,-1}-4\sqrt{\frac{2}{35}}Y_{3,-1}\right)+MM_{221}\left(-\frac{3\sqrt{5}}{5}Y_{1,1}+4\sqrt{\frac{2}{35}}Y_{3,1}\right) \nonumber\\
&& \phantom{XXXX}+\frac{2\sqrt{7}}{7}M\overline{M}_{222}Y_{3,-2}+\frac{2\sqrt{7}}{7}MM_{222}Y_{32}\Bigg)r^4+O(r^5), \\
&&  B_{\theta\theta}=\O(r^5),\\
&&  B_{\theta\varphi}=\O(r^5), \\
&&  B_{\varphi\varphi}=\O(r^5). \label{expansion:cotton2}
\end{eqnarray}
\end{subequations}

In particular, if one has an axially symmetric stationary spacetime
then $M_{221}=M_{222}=0$ and one obtains the \emph{obstruction}:
\begin{equation}
\label{obstruction}
\Upsilon=45S_{110}^2+2\sqrt{5\pi}MM_{220},
\end{equation}
which can be shown to be different from zero for the Kerr solution. Indeed,
\begin{equation}
\Upsilon_{Kerr}=-112\pi M^2a^2.
\end{equation}
Note that $\Upsilon$ is a quantity of quadrupolar nature which closely
resembles the structure of the Newman-Penrose constants of stationary
spacetimes ---see \cite{NewPen68}. However, at this point it is not
possible to exhibit a connection ---if any.

\medskip
\textbf{Remark.} Guided by the suspicion that the Kerr spacetime
admits no conformally flat slices, it would be natural to reckon that
the angular momentum is responsible for this feature. However, Garat
\& Price's perturbative analysis already hints that the responsible of
the non-existence of conformally flat slices in stationary solutions
has to be an object of quadrupolar nature. The quantity $M^2a^2$ is
the only quadrupolar object one can form out of the parameters of the
Kerr solution. Due to the specialness of the latter spacetime, the
form of the obstructions for more general stationary spacetimes cannot
be inferred from merely looking at the Kerrian case.

\section{The Cotton-York tensor in an arbitrary slice}
Before calculating the Cotton-York tensor of the
$\overline{t}=constant$ slices, one has to make some assumptions on the
function $F=F(r,\theta,\varphi)$ defining the coordinate
transformation (\ref{new_slice}). Our Ansatz will be
\begin{equation}
\label{F:Ansatz}
F(r,\theta,\varphi)=F_1(\theta,\varphi)r+\O^4(r^2),
\end{equation}
where the symbol $\O^4(r^2)$ means we assume that $\partial_i F
=F_1+\O(r)$, $\partial_i\partial_j F =O(1)$,
$\partial_i\partial_j\partial_k F=\O(1)$ and
$\partial_i\partial_j\partial_k\partial_l F=\O(1)$. Furthermore, we will
require the metric $\overline{h}_{ij}$, given by equation
(\ref{metric_bar}) to have a well defined conformal compactification
at infinity. This requires the existence of a conformal factor
$\overline{\Omega}$ which is $C^2$ near $i$, such that
\begin{equation}
\overline{\Omega}(i)=0, \qquad \overline{D}_i\overline{\Omega}(i)=0, \qquad \overline{D}_i\overline{D}_j\Omega(i)=2h_{ij}(i),
\end{equation}
where $\overline{D}$ is the covariant derivative with respect to the
metric $\overline{h}_{ij}$. The latter forces $\overline{h}_{ij}$ to
be at least $C^1$ in a neighbourhood of $i$. Now, a short calculation
using Cartesian coordinates and arguments similar to that used in the
remarks to theorem 2.6 in reference \cite{Dai01b} shows that if the
coefficient $F_1(\theta,\varphi)$ contains $l=2$ harmonics then the
metric $\overline{h}_{ij}$ is at best of class $C^{0,\alpha}$ in a
neighbourhood of $i$ \footnote{$f\in C^{p,\alpha}$ means that the
function has $p$-th order derivatives f which are H\"older
continuous with exponent $\alpha$. A function $f$ is said to be
H\"older continuous with exponent $\alpha$ at a point $x_0$ if there
is a constant $C$ such that $|f(x)-f(x_0)|\leq |x-x_0|^\alpha$,
$0<\alpha<1$ for $x$ in a neighbourhood of $x_0$.}. Indeed, the
spherical harmonics $Y_{2m}$ are in Cartesian coordinates\footnote{The
spherical coordinates are singular at $r=0$. Consequently, the
discussion regarding the regularity has to be carried out in Cartesian
coordinates.} given by
\[
Y_{20}=\frac{1}{4}\sqrt{\frac{5}{\pi}}\left(\frac{3z^2}{r^2}-1\right), \quad Y_{21}=-\frac{1}{2}\sqrt{\frac{15}{2\pi}}\frac{z(x+iy)}{r^2}, \quad Y_{22}=\frac{1}{4}\sqrt{\frac{15}{2\pi}}\frac{(x+iy)^2}{r^2}.
\]
Thus, a function $F_1$ made up of these harmonics would be such that
$rF_1=f_1/r+f_2r$ where $f_1$ and $f_2$ are analytic. From here a
``book-keeping'' argument shows that
$\overline{h}_{ij}=\overline{h}^1_{ij}+r\overline{h}^2_{ij}$, where
$\overline{h}^1_{ij}$ and $\overline{h}^2_{ij}$ are analytic and $r\in
C^{0,\alpha}$. The situation is even worse for higher harmonics. On
the other hand, if $F_1(\theta,\varphi)$ contains only $l=0$ and $l=1$
harmonics then $\overline{h}_{ij}$ is at least of class
$C^{2,\alpha}$.

\medskip
Under our assumptions the leading terms of the 3-metric
$\overline{h}_{ij}$ read:
\begin{subequations}
\begin{eqnarray}
&& \overline{h}_{rr}=1+\O(r^2), \\
&& \overline{h}_{r\theta}=-F_1\partial_\theta F_1r^5+\O(r^6), \\
&& \overline{h}_{r\varphi}=-F_1\left(\partial_\varphi F_1-S_{110}\sqrt{\frac{3}{\pi}}\sin^2\theta \right)r^5+\O(r^5), \\
&& \overline{h}_{\theta\theta}=r^2+\O(r^4), \\
&& \overline{h}_{\theta\varphi}=\O(r^6), \\
&& \overline{h}_{\varphi\varphi}=\sin^2\theta r^2+\O(r^4).
\end{eqnarray}
\end{subequations}

The Cotton-York tensor in this case is such that
\begin{eqnarray}
&&\overline{B}_{rr}=-\frac{1}{2\sqrt{\pi}\sin^4\theta}\big( A \partial_{\theta\theta\theta} F_1+ B\partial_{\varphi\varphi\varphi} F_1 + C\partial_{\theta\theta\varphi} F_1+D\partial_{\theta\varphi\varphi} F_1 \nonumber\\
&&\phantom{XXXXXXXXXXXX}+ E\partial_{\theta\theta} F_1+ G\partial_{\varphi\varphi} F_1+ H\partial_{\theta\varphi} F_1+ I \partial_{\theta} F_1 \big)r^3+O(r^4), \label{F_condition}
\end{eqnarray}
where
\begin{eqnarray*}
&& A= -\sin^4\theta\left(4\sqrt{\pi}\partial_\varphi F_1+3\sqrt{3}S_{110}\right), \\
&& B= 4\sqrt{\pi}\sin^2\theta \partial_\theta F_1, \\
&& C= 4\sqrt{\pi}\sin^4\theta \partial_\theta F_1, \\
&& D= -\sin^2\theta\left(4\sqrt{\pi}\partial_\varphi F_1+3\sqrt{3}S_{110}\right), \\
&& E= -\cos\theta\sin^3\theta\left(4\sqrt{\pi}\partial_\varphi F_1+15\sqrt{3}S_{110}\right), \\
&& G= -2\sin\theta\cos\theta\left(3\sqrt{3}S_{110}\sin^2\theta-4\sqrt{\pi}\partial_\varphi F_1\right), \\
&& H= 4\sqrt{\pi}\cos\theta\sin^3\theta\partial_\theta F_1, \\
&& I= \sin^2\theta \partial_\theta F_1 \left( 18\sqrt{3}S_{110}\cos^4\theta -27\sqrt{3}\cos^2\theta +9\sqrt{3}S_{110}+4\sqrt{\pi}\partial_\varphi F_1\right). 
\end{eqnarray*}
Furthermore, one also has that
\begin{equation}
\overline{B}_{r\theta}=\O(r^4),\qquad \overline{B}_{r\varphi}=\O(r^4), \qquad \overline{B}_{\theta\theta}=\O(r^5), \qquad \overline{B}_{\theta\varphi}=\O(r^5), \qquad \overline{B}_{\varphi\varphi}=\O(r^5).
\end{equation}

In order to analyse the condition imposed by the leading term of
$\overline{B}_{rr}$, and consistently with our Ansatz for $F$, we
write
\begin{equation}
F_1=f_{00}Y_{00} + f_{1,-1}Y_{1,-1}+ f_{1,0}Y_{10} +f_{11}Y_{11},
\end{equation}
where $f_{00}$, $f_{1,-1}$, $f_{1,0}$ and $f_{11}$ are complex numbers
such that the coefficient $F_1$ is a real function. The expression one
obtains from substituting the latter into the leading term of
(\ref{F_condition}) ---which we shall denote by
$\Upsilon_F$ is quite complicated. In order to extract its content, we
shall calculate its inner product with different spherical
harmonics. A computation using {\tt Maple V} reveals that
\begin{equation}
\int_0^{2\pi}\int_0^{\pi} \Upsilon_F Y_{00} \sin\theta d\theta d\varphi=-\frac{225}{64}\pi f_{1,0} S_{110}.
\end{equation}
Similarly, one has,
\begin{equation}
\int_0^{2\pi}\int_0^{\pi} \Upsilon_F Y_{2,1} \sin\theta d\theta d\varphi=-\frac{189}{256}\sqrt{15}\pi f_{1,-1} S_{110}.
\end{equation}
Thus, if $\Upsilon_F$ is to vanish then necessarily
\begin{equation}
f_{1,-1}=f_{1,0}=f_{1,1}=0,
\end{equation}
unless, of course, $S_{110}=0$. Now, if $F_1=f_{00}Y_{00}$ then it
follows that the leading terms of the expansions of the Cotton-York
tensor near infinity are identical to those in the expansions given in
the preceding section ---equations
(\ref{expansion:cotton1})-(\ref{expansion:cotton2}). Thus, the (real)
coefficient $f_{00}$ cannot be used to enforce conformal flatness.

We summarise the results of this and the previous sections in the following

\begin{main}
For an asymptotically flat stationary spacetime, necessary conditions
for the existence of a conformally flat slice which in a neighbourhood
of spatial infinity is given by
\[
\overline{t}=t+F(r,\theta,\varphi)=constant,
\]
where $F(r,\theta,\varphi)$ is at least $C^4$ in a neighbourhood of $r=0$, and  can be expanded as
\[
F(r,\theta,\varphi)=rF_1(\theta,\varphi)+\O^4(r^2),
\]
are
\[
M_{222}=M_{221}=45S_{110}^2+2\sqrt{5\pi}MM_{220}=0.
\]
In particular, the Kerr spacetime (with nonvanishing angular momentum)
admits none of such slices.
\end{main}

\section{On expansions at higher orders}

In the light of our main theorem it is natural to ask what happens in
the expansions at higher orders if the conditions given in the main
theorem are fulfilled. In order to keep the complexity of the
calculations at bay, we shall assume that the stationary spacetime is
also axially symmetric. If $45S_{110}^2+2\sqrt{5\pi}MM_{220}=0$ then
for the canonical foliation, the leading term of $B_{r\varphi}$
---which is of order $r^5$--- implies that in order to have a conformal
flatness one needs
\begin{equation}
\sqrt{7\pi}M_{330}\cos^3\theta + 21 S_{110}^2\cos^2\theta-\sqrt{7\pi}M_{330}\cos\theta-21S^2_{110}=0.
\end{equation}
From the last condition it readily follows that
\begin{equation}
M_{330}=0, \qquad S_{110}=0.
\end{equation}
This last result brings further support to the opinion that the only
stationary solutions admitting conformally flat slices are the
Schwarzschild and Minkowski spacetimes.

\section*{Acknowledgments}
I would like to thank R. Beig, W. Simon, S. Dain and H. Friedrich for
interesting discussions and suggestions. I also grateful to R. Lazkoz
and C.M. Losert for a careful reading of the manuscript. This research
is funded by the Austrian FWF (Fonds zur Forderung der
Wissenschaftlichen Forschung) via a Lise Meitner fellowship (M690-N02
and M814-N02). I want to thank the Instituto de Ciencias Nucleares,
UNAM, Mexico, where part of this research was carried, for their
hospitality.


\end{document}